\documentclass{iau}

\usepackage{amsmath}

\def\aj{\textit{AJ}}
\def\apj{\textit{ApJ}}
\def\apjl{\textit{ApJL}}
\def\araa{\textit{ARAA}}
\def\icarus{\textit{Icarus}}
\def\jgr{\textit{Journal of Geophysical Research}}
\def\mnras{\textit{MNRAS}}
\def\na{\textit{New Astron.}}
\def\nat{\textit{Nature}}
\def\pasp{\textit{PASP}}

\def\day{\mathrm{day}}

\def\kyr{\mathrm{kyr}}
\def\Myr{\mathrm{Myr}}
\def\AU{\mathrm{AU}}

\def\covRM{\mathrm{cov}}
\def\inRM{\mathrm{in}}
\def\outRM{\mathrm{out}}
\def\colRM{\mathrm{coll}}
\def\ACov{A_{\covRM}}
\def\AIn{A_{\inRM}}
\def\AOut{A_{\outRM}}
\def\AColl{A_{\colRM}}
\def\OmegaIn{\Omega_{\inRM}}
\def\OmegaOut{\Omega_{\outRM}}
\def\OmegaStar{\Omega_{\star}}
\def\LStar{L_{\star}}
\def\dLStar{\Delta\LStar}
\def\IStar{I_{\star}}
\def\IOut{I_{\outRM}}
\def\RStar{R_{\star}}
\def\RSun{R_{\odot}}
\def\dStar{d_{\star}}
\def\dSun{d_{\odot}}
\def\dEarth{d_{\oplus}}
\def\vRel{v_{\mathrm{rel}}}
\def\tColl{t_{\colRM}}
\def\tOrb{t_{\mathrm{orb}}}

\begin{document}

\lefttitle{Lacki}
\righttitle{Dust to Dust: Prospects for Passive Technosignatures as Relics of ETI}

\jnlPage{?}{?}
\jnlDoiYr{2026}
\doival{10.1017/xxxxx}

\aopheadtitle{Proceedings IAU Symposium}
\editors{J. Haqq-Misra \&  R. Kopparapu, eds.}

\title{Dust to Dust: Prospects for Passive Technosignatures as Relics of ETI}

\author{Brian C. Lacki}
\affiliation{Breakthrough Listen, Department of Physics, Oxford University, \email{astrobrianlacki@gmail.com}}

\begin{abstract}
Technological societies are separated in time, not just space -- that is the lesson of the Drake equation. Might the best way to seek them be to find technosignatures that persist long after their creators? I present work I and my collaborators have done on the idea of passive technosignatures, requiring no upkeep from an active society. These range from microscopic to galactic in scale, including specular reflections from shiny artifacts in the Solar System, lens flares from X-ray binaries, and the survivability of Dyson swarms. I discuss prospects for detecting these technosignatures. In the end, what we may be left with are the end products of collisional cascades: dust.
\end{abstract}


\maketitle

\section{Introduction: transitory intelligence in an ancient universe}
If an alien visited the Earth at a random point in its history, would they observe a technological society? We can't know what will happen in the billions of years to come, but if they visited in the past, the answer is no. Earth has hosted radio broadcasting for $10^2$ years, agriculture for $10^4$, and behaviorally modern \emph{H. sapiens} for $10^5$, give or take. The entire Phanerozoic spans merely about an eighth of the Earth's age. Our visitor usually misses not only us, but the whole reign of complex plants and animals. We are separated not just in space but in time. 

This temporal gulf is implicit in the Drake equation, which includes a slew of factors describing how many stars -- thus, how far in space -- we need to search to find one that \emph{has ever or will ever} give rise to an ETI. The final term is a mean longevity that determines the fraction of societies overlapping us in time. If ETIs are fleeting, then even in standard ``optimistic'' scenarios, our nearest contemporaries could be across the Galaxy. Yet we might find evidence of ETIs around the nearest stars, if only they left a lasting sign (cf. \citealt{Freeman75}). But how could they possibly bridge that gulf in time?

At least one complex system has survived for billions of years: the Earth's biosphere. The ability of living things to replicate enables this persistence,\footnote{\cite{Yokoo79} and others have suggested ETIs could create a kind of passive self-replicating message encoded in genes. The reliability of this method over geological epochs remains unproven.} bridging space and time even as organisms and species come and go. ETIs too might replicate by going out to the stars and establishing new communities. Taken to its natural conclusion, the replication populates the Galaxy over millions of years. If society fails around one star, migrants from another take its place \citep{Wright14}. We have seen no signs of anything like this: this is the heart of the Fermi Paradox \citep{Cirkovic18}. Let us suppose some still-unknown factor has suppressed a Galactic ecology. How else might someone make their existence known across cosmic time?

We need a relic technosignature: something recognizably artificial that manages to survive far longer than any of our technology without any possible maintenance or upkeep (as in \citealt{Balbi21}). Solar System artifacts are a strong contender. Interplanetary space and inactive planetary bodies like the Moon lack weathering, the only hazards being occasional meteoroid impacts and radiation exposure. Relics in these environments could last millions of years, far outliving the ETIs themselves \citep{Rose04,Davies12,Benford21}. But if no artifact reaches our Solar System specifically, we need to look for some kind of radiative technosignature. Even locally deposited artifacts might use broadcasts to get our attention in the vastness of interplanetary space \citep{Bracewell60}. Could they set up a passive broadcasts, emitted without any effort or artificial power?

\section{The fundamentals of passive broadcasts}
Beacons are classic technosignatures. They emit broadcasts meant to be easily detected and recognized as artificial, regardless of information content. Active technological broadcasts require large power plants, control mechanisms, and so on, in turn implying active maintenance when they break down. An abandoned beacon cannot rely on that. Fortunately, the Universe already has plenty of radiating objects like stars that can act as signal lamps. The passive beacon uses a \emph{modulator} to create artificial fluctuations in its lamp's luminosity. These can create positive fluctuations in the lamp's apparent brightness (making it appear brighter) or negative (apparently dimming it); the behavior may even vary with time and frequency.

Consider an idealized modulator orbiting a lamp with surface brightness (surface flux per unit solid angle) $\IStar$ and radius $\RStar$ at a distance $\dStar$. Lamp-light coming from a solid angle $\OmegaIn = \OmegaStar \approx \pi (\RStar/\dStar)^2$ enters the modulator through an input area $\AIn$. A fraction $\eta$ of that power is emitted through an output of area $\AOut$ into a cone of solid angle $\OmegaOut$. The modulator may also cover an area $\ACov$ of the star. Working in a spectral band that excludes the modulator's waste heat, the apparent change in the lamp's luminosity is, by conservation of energy,
\begin{equation}
\label{eqn:PassiveLuminosity}
\dLStar = 4 \pi [\IOut \AOut - \IStar \ACov] = 4 \pi \IStar [\eta \omega \AIn \OmegaIn/\OmegaOut - \ACov] ,
\end{equation}
where $\IOut$ is the observed surface brightness of the modulator and $\omega$ is $1$ if the observer is in $\OmegaOut$ and $0$ outside of it.\footnote{I exclude waste heat because typically the instruments that observe the modulation cannot detect it (e.g., optical CCDs are not sensitive to mid-infrared). When included, waste heat is relatively faint because, like the output of a diffuser, it is emitted nearly isotropically. A modulator is cooler than its lamp.} While light going into a small $\AIn$ can be beamed out of large $\AOut$ under etendue conservation (a reversed telescope), modulators can much simpler -- a single element like a metal foil, lens, or flat mirror suffices, with $\AOut = \AIn$ and $\OmegaOut = \OmegaIn$.

A modulator that simply scatter light nearly isotropically ($\OmegaOut \sim 4 \pi$) is here called a \emph{diffuser}. A diffuser can serve as a passive beacon: for example, it might reflect only an unusual color or circularly polarized light. While the ``broadcast'' is visible to nearly anyone, it is faint \emph{because} its light goes in all directions, with $\dLStar \sim \pi \IStar A (R_{\star}/d_{\star})^2$. Equivalently, $\IOut$ is small because the diffuser averages the bright lamp with the dark sky around it. Indeed, non-transiting terrestrial exoplanets are diffusers and their light is difficult to detect.

A bright passive beacon maximizes surface brightness. The Second Law of Thermodynamics sets a hard limit for a passive optical system: surface brightness is at best conserved ($\IOut \le \IStar$). A \emph{glinter} is a perfect mirror or lens that reaches the limit when properly aligned, achieving $\dLStar = +4 \pi \IStar A$. An \emph{occulter} is a modulator that simply blocks out part of the star, so that $\dLStar = -4 \pi \IStar A$. Glinters and occulters are much easier to detect than diffusers, but the signal is beamed into $\OmegaStar$. This fundamental trade-off follows directly from energy conservation -- high surface brightness implies rare events.

\section{Glints from relic mirrors: passive beacons in the Solar System}
Let's first consider a passive beacon in the Solar System. If ETIs have left behind an artifact in space for us, a passive beacon can help us find it. A free-floating planar mirror (or a flat, specularly reflective surface like a solar panel acting like one) is a glinter, reflecting sunlight into a cone opening into angular radius $\theta_{\odot} \approx \RSun / \dSun = 0.27^{\circ} (\dSun / \AU)^{-1}$. From our perspective, the mirror usually reflects empty space and is completely dark, but when the cone sweeps over us, we see a \emph{glint}. Now, a relic mirror with no attitude control cannot aim. Instead, the cone sweeps past the Earth at a speed $\vRel$ because of orbital motion and the glint lasts for
\begin{equation}
\tau_{\mathrm{glint}} \sim \theta_{\odot} \dEarth / \vRel = 0.81\ \day ({\dEarth}/{\dSun}) ({\vRel}/{10\ \mathrm{km~s}^{-1}}).
\end{equation}

Glints are a proven technosignature. We observe them from our own satellites in Earth orbit, so often they are a kind of ``RFI'' for rapid optical transient searches \citep{Corbett20}.\footnote{\cite{Villarroel22} aims to suppress this foreground by examining archival pre-Space Age photographs. Subsequent claimed detections, and whether they might be spurious, are still debated.} Anthropic glints come from Earth orbit, however, where they last only seconds. Relic mirrors in interplanetary space would distinctly glint for several hours. A non-spinning mirror as small as 100 cm$^2$ is visible at $\dEarth = 1\ \AU$ in the LSST \citep{Lacki19}. However, the probability any one randomly mirror is visible in a single snapshot of the entire sky is $\sim \Omega_{\odot} / (4 \pi) \sim 5 \times 10^{-6}$. 

Relic mirrors lacking attitude control may be spinning. Rotation causes the cone of reflected light to sweep a band in the mirror's sky over one rotation period. From our perspective, the glint resolves into a series of pulses with low duty cycle. Averaging over phase, the glint is much dimmer but we are more likely to see it. The tradeoffs are somewhat complicated, but while slow spinning may enhance the effective sample volume, \cite{Lacki19} finds that very rapidly spinning mirrors are much harder to detect: $A \gtrsim 30$ m$^2$ is visible in the LSST at 1 AU.

\section{Considerations for interstellar passive beacons}
In the Solar System, glinters are clearly superior to occulters. A glint occurs against the dark night sky, where the backgrounds are small; an occulter transits the Sun, with an incredibly high shot noise and natural surface variations. That's why we do not observe the transits of natural meter-scale meteoroids across the Sun. Interstellar passive beacons cannot be resolved from the lamp, and thus both glinters and occulters fight against the lamp's shot noise. Thus small glinters and occulters with the same area have the same detectability -- the blip from a transiting rough metal sheet is just as big as one from a finely-machined lens. The only difference is that a glinter \emph{adds} an area glowing with $\IStar$, while the occulter \emph{subtracts} that area. 

\cite{Arnold05} first proposed occulters as interstellar passive beacons. Because there is a foreground of natural occulters like transiting exoplanets, Arnold suggested that they would have unusual shapes we would infer from the light curve. Searching for transits with unusual features is now established in SETI \citep{Zuckerman24}. The ``inverted transit'' of a glinter is also likewise anomalous. Although a glinter event is only as detectable as an occultation, it may be more \emph{recognizable} because its anomalous characteristic is the brightening itself. But a single occulter is a poor beacon. The signal from each modulator sweeps out a band of width $2 \theta_{\star}$ in its sky as it orbits (analogous to our rotating mirror in the previous section), visible from a small range of inclinations. For an ``isotropic'' beacon, the number of modulators is:
\begin{equation}
N_E \gtrsim \pi d_{\star}/(2 R_{\star}) \approx 340 (d_{\star}/\mathrm{AU}) (R_{\star}/R_{\odot}) .
\end{equation} 

Glinters may have another upside: there is no fundamental limit to their luminosity. While an occulter can only block out 100\% of a lamp, a glinter could hypothetically be much larger than the lamp itself ($\dLStar/\LStar = \AOut / A_{\star}$). For a sun, it essentially requires a Dyson swarm of mirrors, all reflecting sunlight in the same direction. This implies attitude active control, not a passive beacon. The challenge is finding a lamp small enough a single structure could dwarf it, but bright enough to be seen at galactic distances.

\section{Lens flare: intergalactic passive broadcasts?}
A superluminous glinter needs a hot lamp to maximize surface brightness. The hottest thermal sources we know occur in Low Mass X-ray Binary (LMXB) systems.\footnote{
Nonthermal emission can reach even higher brightness temperatures. A radio-emission zone on a pulsar may also be a good lamp. \cite{Chennamangalam15} considers occulters, but in principle there could be radio lens flares, possibly resembling fast radio bursts \citep{Lacki20}. Natural analogs from plasma lensing have been observed in certain pulsar systems \citep{Main18}.} Material reaches near relativistic orbital speeds as it accretes onto a compact object, but if the object is a neutron star, it must come to a complete stop almost instantaneously, releasing all of the kinetic energy at once (e.g., \citealt{Popham01}). The impact region is the boundary layer, a mere few square kilometers with a luminosity of up to $\sim 10^5 L_{\odot}$ -- an ideal lamp. 

LMXBs are enticing lamps for passive beacons, \emph{if} ETIs can reach them. The potential for occulters has already been noted \citep{Imara18}, but we can go further with a glinter larger than the boundary layer. Consider a collimating lens, magnifying whatever is directly behind its center. Most of the time, it magnifies the much colder accretion disk or empty space. Indeed, an off-center lens much larger than our lamp acts as an occulter. If the lens center passes directly in front of the boundary layer, the entire lens lights up with its surface brightness. A lens flare results. A 1,000 km radius lens can appear as luminous as a billion suns for a fraction of a second -- maybe bright enough for an \emph{intergalactic} beacon.

The catch is that this luminosity is mostly hard X-rays. Both refractive and diffractive hard X-ray optics suffer from chromatic aberration, thus focusing only in a narrow range of energies and diluting the lens flare signal. The lenses also need some passive method to point themselves at the LMXB. Our own ability to detect X-ray transients is limited; we have no X-ray equivalent of the Rubin telescope. \cite{Lacki20} shows that an $R =$ 1,000 km achromatic lens flare could be detectable in M31 with our most sensitive X-ray facilities. But we almost certainly would miss them, because they are rare events. The boundary layer is a small moving target and the swarm must be located far from the binary, tens of AU, to not instantly sublimate from the intense luminosity. Indeed, there may not be enough local solid matter to build a lens swarm in the first place -- a very large number of lenses is needed to ensure a decent flare rate. For now, if these lens swarms exist, we probably will only find them in our own Galaxy and possibly the Magellanic Clouds, where they would be bright enough to be picked up by widefield monitors.

\section{The mortality of megaswarms: collisional cascades}
A megastructure is a technological artifact of astronomical scale. The archetypal megastructure is a Dyson sphere, which surrounds a sun and has covering factor $\sim 1$.\footnote{All Dyson spheres are thus occulters. In addition to their positive waste heat, itself a passive broadcast, the abnormal dimming of their host star is a potential technosignature \citep{Zackrisson18}.} In general, a megastructure is envisioned as a megaswarm, with a large number of elements in orbit around a host like a star; no single solid structure of that scale could survive \citep{Wright20}. Interstellar passive beacons, including occulter swarms and X-ray lens swarms, are sparse megaswarms, with each sightline uncovered most of the time. Both Dyson spheres and passive beacons need elements with a full range of orbital inclinations to intercept all possible sightlines.\footnote{\cite{McInnes26} has an elegant solution: radiation-pressure supported elements can simply float at high latitude. Degradation in attitude control or reflectivity might upset the balance needed for this configuration.}

The scale of a megaswarm and the effort to build one is imposing. But if the elements of a swarm are unguided and their orbits are even partially randomized, they collide with each other. A collisional cascade ensues, the hypervelocity fragments from each collision obliterating further elements, initiating a rapid grinding process that smashes everything into dust. Collisional cascades are inferred to happen for minor body populations throughout the Solar System and in the debris disks of exoplanetary systems (e.g., \citealt{Wyatt08,Hughes18}). Technological collisional cascade is also a threat to the growing swarm of satellites in Earth orbit \citep{Kessler78}. Our satellites take evasive actions, expending propellant. A long abandoned megaswarm, where the elements have lost maneuvering abilities and lack self-repair, has no way to avoid this fate \citep{Lacki25}.

The collisional time describes the instantaneous rate an element can expect to be hit by another in a fully randomized swarm. For a naive estimate, suppose the megaswarm fills a shell around a host with radius $R_S$ and thickness $r_S$. The swarm has $N_E$ elements placed on random orbits, with periods $\tOrb$. Typical relative velocities $\vRel$ are comparable to the circular orbital velocity. Likewise, the collisional cross sectional area $\AColl$ is of order the projected area of each element $A$. Defining $n_E$ to be the number density of elements, the collisional time is
\begin{equation}
\tColl = (n_E \AColl \vRel)^{-1} = 4 \pi R_S^2 r_S / (N_E A_{\mathrm{coll}} \vRel) \sim \tOrb \cdot F_S^{-1} \cdot (r_S / R_S),
\end{equation}
where $F_S = N_E A /(4 \pi R_S^2)$ is the covering fraction of the megaswarm. Thus a filled megaswarm like a Dyson swarm, left to its own devices and with randomized orbits, would immediately self-destruct. Even sparse megaswarms have a limited life: $\tColl \sim 1\ \Myr$ for occulters around a solar twin at 1 AU. A lens swarm around an LMXB is estimated to have $\tColl \sim 100\ \kyr$. While long, the lifespans are mere stubs in geological time.

This estimate proves robust even when the orbits are not fully randomized. The underlying issue is the wide range of inclinations. To impose order on the megaswarm, the elements could be confined to concentric circular belts, each a small range of semimajor axis and inclination. The small mutual inclinations reduces the relative velocities, but at the same time, it reduces the fraction of the shell volume that is occupied -- all of the space at high relative inclination is wasted. In fact, relative velocity trades against density so that the collisional time is unchanged. Actual collisional cascades can proceed much faster than $\tColl$ -- any one of the numerous tiny fragments is moving fast enough to shatter a large element.

The cascade begins once the orbits of elements start crossing. Many stellar systems have either a companion star or orbiting planets, either of which are a source of perturbations. The Lidov-Kozai instability is a particular threat to megaswarms with a wide range of inclinations: circular orbits above a threshold inclination oscillate into highly eccentric orbits with low inclination and back. Higher-order effects can operate as well (e.g., \citealt{Naoz13}). The timescale for a full Lidov-Kozai cycle is around 150 kyr for an element at 1 AU perturbed by a Jupiter analog. Even absent gravitational perturbations, elements can drift off-orbit from their own radiative thrust, as small asteroids do \citep{Bottke06}.

A megaswarm has an inherent self-destruction mode, leaving us back where we started: as a relic technosignature, it requires maintenance. A Dyson swarm is not a good monument; and even a passive beacon is transient in the cosmic scheme of things. Any sample of abandoned megaswarms we observe will be biased to those with long lifespans. These could be found around solitary stars with no perturbing companions. When there is a companion, Lidov-Kozai perturbations are slower in swarms far interior or exterior to it -- although swarms close to a star are especially vulnerable to radiatively induced perturbations. Very cold megaswarms, operating far from their star, however, generally can have long lives and thus may dominate the observable population. The ultimate extreme would be artificial elements distributed in the interstellar medium itself, untethered to any star, where the Galactic orbital period is a few hundred million years. \citet{Lacki16} investigated these ``blackboxes'', arguing they could take the form of thin dipole needles to absorb and radiate in microwaves. Instead of being ground to dust, a blackbox is already a form of dust.

\section{The final passive technosignature}
In the absence of easy interstellar travel, short-lived ETIs trying to memorialize themselves may try to build a passive beacon visible from interstellar distances. These could be swarms of modulators around a lamp like a star, but the elements must be big and numerous for maximum signal. In turn, these swarms are susceptible to collisional cascades, eventually smashing each of these elements into bits. Without the ability to replicate and regenerate themselves, it seems any sign of an ETI might vanish. Might there even have been a Kardashev Type III ETI pervading our Milky Way a billion years ago that vanished without a trace?

And yet, in being ground to dust, the final act of a passive beacon is to spread itself among the stars. As the fragments are broken into smaller pieces, the ratio of radiation pressure to gravity rises. Once it exceeds $\frac{1}{2}$ -- typically micron-scale grains for a sunlike star -- the broken pieces are blown out of the system, removed from the cascade and saved from further destruction \citep{Hughes18}. Thus, even if the builders never go to the stars, the fine remnants of their works do, as long as their home sun is bright enough. Each broken swarm releases an astronomical number of grains into the interstellar medium. As the Solar System passes through this reservoir, the worlds within it would be exposed to a sprinkling of technograins, and a small fraction of those can land softly enough in the regolith of worlds like the Moon to maybe be recognizable even over billions of years (cf. \citealt{Arkhipov96}). And so we are back again at the prospect of Solar System artifact by a different route: not macroscopic objects deliberately placed here, but dust, the unintended microscopic testament to our possible predecessors still waiting to be found (Pinault et al. 2026, submitted).

\section*{Acknowledgments}
I thank the Breakthrough Listen program for their support. Funding for Breakthrough Listen research is sponsored by the Breakthrough Prize Foundation (\url{https://breakthroughprize.org/}). I am grateful for the collaboration of Lewis Pinault, Ian Crawford, and Andrew Siemion on micron-scale fragments as technosignatures, and to Jason Wright and several referees for comments and discussion.

\end{document}